\documentclass[runningheads]{llncs}
\usepackage[T1]{fontenc}
\usepackage{graphicx}
\usepackage{color}
\PassOptionsToPackage{hyphens}{url}\usepackage[pdftex, colorlinks=true, hyperfootnotes=false, hyperindex=true, plainpages=false, pagebackref=false, pdfpagelabels=true, pdfstartview=FitH, linkcolor=blue, citecolor=blue, urlcolor=blue, bookmarks=false]{hyperref}

\usepackage{annotate-equations}
\usepackage{amsmath}
\usepackage{amssymb}

\usepackage{url}
\usepackage{booktabs}
\usepackage{multirow}
\usepackage{makecell}
\usepackage[inline]{enumitem}
\usepackage{tabularx}
\usepackage{wrapfig}
\usepackage[subtle, mathspacing=normal]{savetrees}

\usepackage{tikz}
\usetikzlibrary{arrows.meta, calc, petri, positioning}

\begin{document}
\title{An Architecture for Decentralised Deployment and Operation of Blockchain Applications}
\titlerunning{Decentralised Deployment and Operation of Blockchain Applications}
%
\authorrunning{F. Stiehle et al.}
%
\author{Fabian Stiehle\inst{1} \and Kirill Inozemtsev\inst{1} \and Ingo Weber\inst{1,2}}
%
%
%
\institute{
Technical University of Munich, School of CIT, Germany, \email{first.last@tum.de} \and
    Fraunhofer Gesellschaft, Munich, Germany}
%
%
%
%
%
\maketitle              
\begin{abstract}
Blockchains and distributed ledger technologies allow the operation of manifold decentralised applications (dApps). Such applications are based on smart contracts, a programmable abstraction that is executed in a decentralised manner.
To ensure the correctness of smart contracts, blockchain application developers rely on DevOps practices such as automated testing and continuous integration and deployment. However, such infrastructure is often controlled by single entities. 
For larger blockchain applications, this issue is resolved by relying on concepts of Decentralised Autonomous Organisations (DAOs), which allow proposals to be autonomously executed once they reach a pre-defined quorum. Such a governance architecture is complex and requires integration with existing patterns for contract discovery and upgradeability. 
In this paper we integrate these concepts considering DevOps best-practices into a novel architecture that remains agnostic to different governance and upgrade implementations. We extend the known registry pattern to support deterministic deployments and present a decentralised deployment framework, including integration and deployment pipelines, user-interfaces, and version control integration. In our approach, each party implements and verifies their own tests before engaging in the use of a (newly deployed) smart contract. We provide a reference implementation, available as open-source, and evaluate the proposal thoroughly. Our architecture can serve as a reference for future integrations, while our open-source framework is aimed at reducing the complexity of adopting such a process in practice.
\end{abstract}
%
%
%
\section{Introduction}
%
 
Blockchain is well suited as a basis for applications that involve multiple parties with low mutual trust. 
Business logic can be implemented in smart contracts and then independently verified by all parties. The smart contract becomes the new foundation of cooperation, due to its ability to enforce business rules and provide a secure audit trail~\cite{stiehle2022blockchain}. Thus, the code of the deployed smart contract is of central importance---specifically that the code meets expectations in terms of correctness, functionality, and quality attributes.

\textit{DevOps} can be seen as collection of practices concerned with the same aspects: getting new software code into production quickly and in high quality~\cite[Ch. 1]{Bass2015DevOpsBook}. 
It comprises concepts like continuous integration (CI): frequently integrating and merging development work;  continuous delivery: having a production-ready artifact built after each development cycle; and continuous deployment (CD): frequent deployment to the production environment.
An important tool to achieve this is an automated deployment pipeline, covering software building, testing, and deployment~\cite[Ch. 5]{Bass2015DevOpsBook}. 

While blockchain application development comes with its own set of challenges, DevOps concepts are already applied in blockchain application development practice. However, while the goal is to run the application in a decentralised manner, deployment is often centralised: the deployment pipelines run in the environment and control of a single actor~\cite{wohrer2021devops}.
While this might be more permissible in cases where the blockchain application is operated more traditionally, e.g., in an business-to-consumer (B2C) fashion, 
in a multi-party setting, e.g., B2B collaboration, this re-introduces trust issues---which blockchain intended to resolve. 

The evolution of larger blockchain ecosystem applications~\cite{wohrer2021devops}, like layer two Rollups,\footnote{For example, the governance implementation of the \textit{Arbitrum} DAO, \url{https://github.com/ArbitrumFoundation/governance/blob/main/docs/overview.md}, accessed 2026-03-24.} are governed by concepts of Decentralised Autonomous Organisations (DAOs). DAO architectures allow arbitrary proposals to be autonomously executed once they reach a pre-defined quorum. 
However, such a governance architecture is complex and still requires integration with existing patterns for smart contract discovery and upgradeability. 

As we show, while DevOps practices for blockchain applications (e.g.~\cite{wohrer2021devops,gorski2021towards}), upgradeability and upgrade mechanisms (e.g.,~\cite{kumar2025leagan,bodell2023proxy}), and governance (e.g.,~\cite{han2025daoreview,xu2025patternextended}) have all been studied in isolation, no prior work delivers an integrated architecture facilitating the application of DevOps best-practices in a multi-party, decentralised manner. 

The goal of the work presented here is to address this gap.
Given the apparent prevalence of traditional, centralised setups in practice, we see this work as highly relevant~\cite{wohrer2021devops}.
As we show, integrating these concepts is not trivial, and requires adoption and extension of multiple existing patterns of blockchain architecture. Specifically, we introduce the \textit{deterministic registry}, a variant of the contract registry pattern, to facilitate deterministic deployment.

In summary, we present an architecture realising a decentralised deployment pipeline, to enable continuous deployment in a low-trust environment. It allows parties to independently perform various verification tests and conducts an automatic bytecode comparison between the tested and the deployed version of smart contracts. 
The on-chain logic extends the registry pattern and integrates state-of-the-art governance architectures, facilitating a quorum-based voting process. 
We provide a reference implementation and evaluate it in terms of feasibility, correctness, and cost. 
Our contributions are:
\begin{itemize}
    \item A \textbf{reference architecture} integrating governance, upgradeability, and DevOps practices, for blockchain applications, agnostic to specific governance and proxy implementations.
    \item The \textbf{deterministic registry pattern}, extending the known registry pattern with deterministic address derivation to enable cryptographic commitment to deployment artefacts.
    \item An \textbf{evaluation} of the proposed artifacts via feasibility assessment, correctness benchmarking (process conformance checking), and gas-cost analysis,  based on an integrated open-source implementation and accompanied by a replication package.
\end{itemize}

The remainder is structured as follows. Related work is discussed in \autoref{sec:relwork}. The requirements and conceptual solution are presented in \autoref{sec:solution}.
On the basis of the prototype we evaluated our conceptual solution, as described in \autoref{sec:eval},
before \autoref{sec:disc} discusses the work and concludes the paper.

\section{Related Work}\label{sec:relwork}
The work presented in this paper builds on three areas: (i) DevOps practices for blockchain application development, (ii) smart contract upgrade mechanisms, and (iii) decentralised governance. Each of these areas has received attention, as we will discuss in the following review of the state of the art.

Wöhrer and Zdun~\cite{wohrer2021devops} investigate the state of the art and practice in \textbf{DevOps practices and deployment pipelines for blockchain applications}, by surveying grey literature and GitHub repositories. They find that established solutions and tools provide sufficient means to build and test smart contracts. According to their findings, the use of corresponding tools and frameworks for smart contract development is essential, such as \textit{Truffle} (now deprecated) or \textit{Hardhat}, which provide means for compilation, automated testing, and deployment. Static analysis tools and test coverage reports can be used in the test phase. Since smart contracts are immutable, Wöhrer and Zdun propose to implement an upgrade mechanism through the use of the proxy pattern---which is now established practice and manifold proxy implementations exist~\cite{bodell2023proxy}. 
Notably, in their study, they discuss a centralised workflow for CI and CD~\cite{wohrer2021devops}. They do not report of any discovered decentralised approaches. As the deployment of smart contracts to the production environment (possibly the blockchain mainnet) is infrequent and security critical, Wöhrer and Zdun propose a manual approval step (i.e., continuous delivery, not continuous deployment as per Bass et al.~\cite{Bass2015DevOpsBook}).

Shahin et al.~\cite{shahin2017continuous} present a literature survey on the general use of continuous integration, continuous delivery, and continuous deployment. They report that distributed organisations have difficulties introducing these practices due to a lack of visibility, as a consistent perception is difficult to achieve. 
Górski addresses continuous delivery and continuous deployment for blockchain applications~\cite{gorski2021towards,gorski2021continuous}. Both works describe a deployment pipeline capable of building deployment packages for R3 Corda. However, the deployment is centralised and performed from a trusted node.
Similarly, Nasr et al.~\cite{nasr2024hybrid} incorporate a centralised CI/CD pipeline to support the deployment of applications in their blockchain deployment framework. In practice, tools like \textit{OpenZeppelin Defender} or \textit{Tally} provide proprietary centralised platforms to manage deployment or proposal processes.\footnote{Defender, \url{https://docs.openzeppelin.com/defender} (discontinued in June 2025); and Tally, \url{https://www.tally.xyz/}, both accessed 2026-05-19.}

There are also works focusing on \textbf{using blockchain technology to improve DevOps practices} of traditional applications~\cite{saleh2024blockchain,beller2019blockchain,akbar2022toward}, e.g., to improve transparency~\cite{akbar2022toward}. This is orthogonal to our work. Bankar and Shah~\cite{bankar2021blockchain} propose a system of managing DevOps pipelines, where each software update is logged to the blockchain. 
Akbar et al.~\cite{akbar2022toward} propose to use blockchain as a means to store metadata information about developed software. 
In their position paper, Beller and Hejderup~\cite{beller2019blockchain} sketch an integration system, where build jobs are executed by nodes in a blockchain network. 

Additionally, multiple works study \textbf{upgradeability and upgrade mechanisms on blockchain}~\cite{kumar2025leagan,bodell2023proxy,benedetti2024comparative}. To achieve upgradeability for blockchain-based applications, so-called proxy patterns are utilised. A proxy maintains a stable contract address across contract versions; functionality is updated through deploying new contracts. The proxy is then updated to point to the new functionality. Numerous proxy patterns have been observed in practice~\cite{bodell2023proxy}. A proxy implementation can entail multiple smart contracts. For example, the Diamond standard, which can be seen as a proxy utilising a decorator style pattern to compose (upgradeable) functionality from separately deployed contracts~\cite{eip2535}, while providing a stable (proxy) address. 
A proxy implementation can be combined with the registry contract pattern~\cite{xu2018pattern}: a contract maintaining a mapping of name to current (proxy) address of (a projects) smart contracts. 
Additionally, a proxy implementation often incorporates the data contract pattern~\cite{xu2018pattern}, where data storage is separated from functionality to maintain access to data throughout newly deployed contract versions.

Similarly, the problem of \textbf{governance} in multi-stakeholder blockchain application environments is studied extensively. Arguably, this is most well-understood in the context of decentralised autonomous organisations~\cite{han2025daoreview}. In addition, there exist well-known patterns such as the multiple and dynamic authorisation~\cite{xu2018pattern} or the  decentralised oracle pattern~\cite{xu2025patternextended}. In essence, these enable support for transactions that require quorum-like consensus before being executed.

A \textit{de facto} standard for well-audited \textbf{best-practice implementations} of various patterns are the libraries provided by OpenZeppelin.\footnote{\url{https://docs.openzeppelin.com}, accessed 2026-03-24} For example, their Governor module provides a flexible implementation supporting the creation of ballots from arbitrary proposals submitted on-chain.\footnote{\url{https://docs.openzeppelin.com/contracts/5.x/governance}, accessed 2026-03-24} This governance concept can be used in practice to govern the evolution of a blockchain application.\footnote{For example, see the upgrade process of the DAO that governs the evolution of the layer two implementation~\textit{Arbitrum}, \url{https://github.com/ArbitrumFoundation/governance/blob/main/docs/overview.md}, accessed 2026-03-24.}

To the best of our knowledge, this paper presents the first integrated architecture for a decentralised deployment process, integrating existing patterns and DevOps best practices for blockchain applications.
To do so, we extend the known registry pattern to support deterministic deployments, given consensus on a deployment proposal is reached.

\section{Architecture Overview}\label{sec:solution}
Following our research goal \textit{to design an approach for coordinated decentralised deployment} and against the backdrop of the related work, specifically regarding existing recommendations of DevOps practices~\cite{Bass2015DevOpsBook} and their use for blockchain applications~\cite{wohrer2021devops}, we formulate our \textbf{main requirements} R1--R3, as depicted in \autoref{tab:requirements}.
\begin{table}[t]
\centering
\scriptsize
\caption{Main requirements and our architecture response.}
\begin{tabularx}{\textwidth}{l@{\hspace{.3cm}} X@{\hspace{.3cm}} X}
\toprule
\textbf{} & \textbf{Requirement} & \textbf{Architecture Response} \\
\midrule
R1 & \textbf{Decentralisation}: Each stakeholder must be able to test each software version independently, including defining their own (private) tests. 
& Integrated off-chain integration pipelines, including test environments, in which each new version is tested independently. \\
\midrule
R2 & \textbf{Consensus}: No stakeholder has singular control over the software release cycle, potentially releasing software versions that benefit them over others. Once there is consensus on a new version, this consensus can not be repudiated. 
& Integrated governance smart contract, facilitating on-chain voting on each new version. Deployment is only possible if a positive quorum on a new version is reached. Votes are recorded on-chain, ensuring non-repudiation.\\
\midrule
R3 & \textbf{Version Authenticity}: Each stakeholder must have the means of verifying that the deployed version is the same as the version they tested locally. 
& Cryptographic assurance of deployments through the deterministic registry, which binds a version's bytecode to its deployment address.\\
\bottomrule
\end{tabularx}
\label{tab:requirements}
\end{table}
\begin{figure}[t]
    \centering
    \includegraphics[width=1\linewidth]{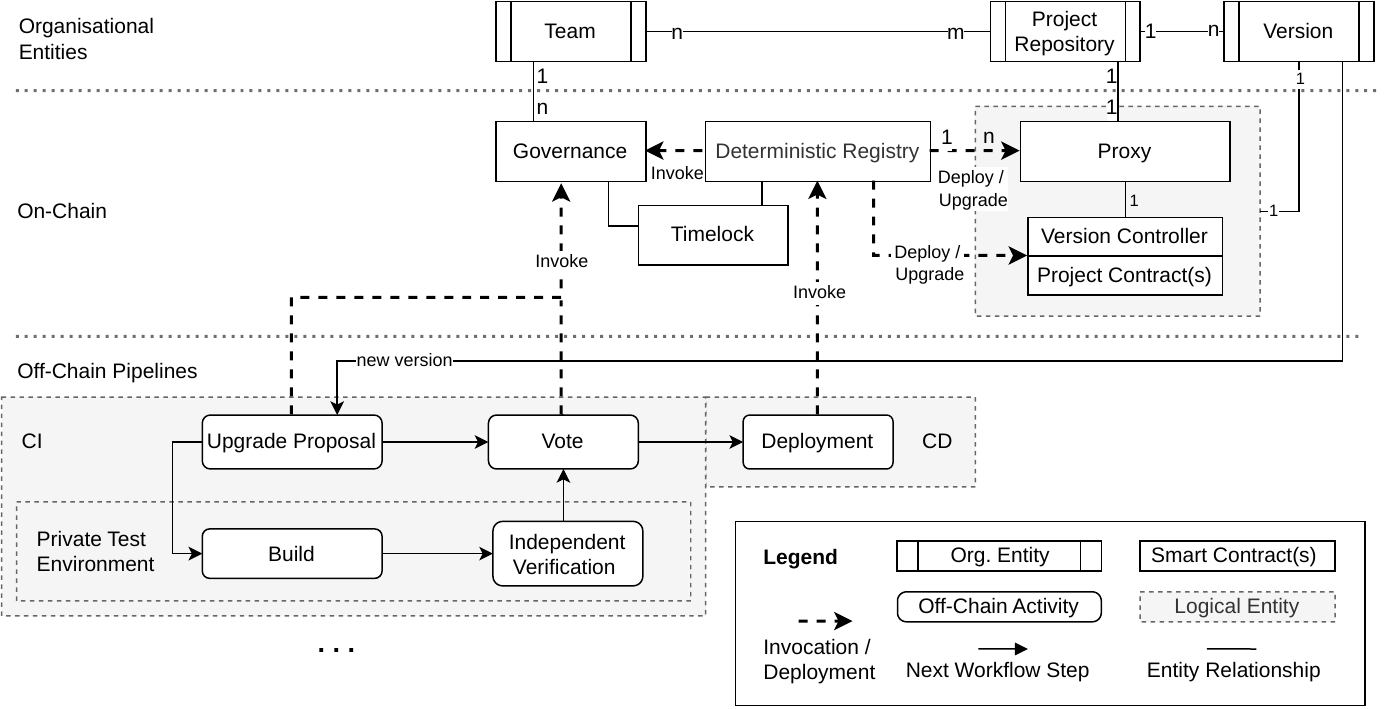}
    \caption{Overview of the proposed architecture with on- and off-chain entities.}
    \label{fig:overview}
\end{figure}
The requirements of traditional deployment pipelines apply in addition~\cite[Ch.\ 5]{Bass2015DevOpsBook}. Examples include traceability (which version of the code and configuration was used in a given run of the pipeline), integrity (a.o.\ preventing the unauthorised insertion of malicious code), and quality attributes related to usability (easy-to-use tooling and graphical interfaces), and automation (integration with state-of-the-art tools, e.g., GitHub Actions).

We propose an architecture that fits the requirements as outlined in \autoref{tab:requirements}. \autoref{fig:overview} shows a high-level overview. The main intuition of our construction involves each stakeholder deploying an independent integration pipeline, in which each new version is tested independently, achieving R1. An on-chain registry serves as single authority of currently deployed contract versions. To achieve R2, it makes use of a governance component, which facilitates on-chain voting on each new version. To achieve R3, we extend the registry pattern to allow deterministic deployments. We define deterministic deployments as being deterministic w.r.t. the bytecode and resulting smart contract addresses the contracts are deployed to. Thus, the contract address effectively becomes a cryptographic commitment to the included bytecode (and vice-versa).

Next, we explain the architecture in more detail and introduce the deterministic registry. Consequently, we outline the proposed CI/CD workflow in more technical detail before discussing (final) deployment and governance.

We consider the case where a project has the following organisational entities (upper row of \autoref{fig:overview}): A version control system (\textit{project repository}) hosts (likely multiple) smart contract source files, tests, and other scripts. The repository is maintained by a group of stakeholders with different roles, forming a \textit{team}.

The on-chain components (middle row of \autoref{fig:overview}) consist of the following smart contracts. The \textit{Governance} contract ensures the decentralised governance (voting on new proposals). The \textit{Deterministic Registry}, an extension of the registry pattern, handles the deterministic deployments of new versions and maintains a registry of current versions and their corresponding blockchain addresses.
To increase upgradeability, \textit{Proxy} contracts can implement the data contract pattern to separate functionality from storage. A proxy maintains the current address of the current approved version of a project's smart contract(s).\footnote{One version can entail multiple proxy contracts, which correspond to different (data) instances using the implementation; See, for example OpenZeppelin's \textit{BeaconProxy} implementation; \url{https://docs.openzeppelin.com/contracts/5.x/api/proxy}, accessed 2026-03-24.} 

The registry is agnostic to different proxy and governance implementations, as they depend on context-specific factors. A diverse range of instantiations exist in practice (c.f.,~\cite{bodell2023proxy} for proxies and \cite{han2025daoreview} for governance). Thus, our architecture is intentionally kept agnostic to specific proxy and governance implementations. 

If a project consists of multiple contracts (which we anticipate), one contract must serve as the \textit{Version Controller}, the main entry point for the application, to support their deterministic deployment.
An optional \textit{Timelock} contract ensures that each stakeholder has time to react to an imminent accepted contract upgrade by allowing proposals to be queued for a specified time before they are deployed.

Off-Chain (bottom row of \autoref{fig:overview}), each voting stakeholder deploys a local off-chain  pipeline to facilitate CI and CD. These pipelines implement the following workflow. Once a new merge ('pull') request to the protected main branch of the shared repository has been made (therefore proposing a new version), the proposer of the request creates a new \textit{Upgrade Proposal}, which attaches an identifier to a given state of the source code (e.g., a commit hash). To do so, the proposer invokes the governance contract. All stakeholders granted voting rights can vote on the upgrade until some predefined quorum or timeout is reached. To verify the upgrade, stakeholders pull the source code linked to the proposal and verify it using a private test environment, which will include private tests and audits, depending on the deployment target (e.g., test or mainnet).
Once they have assessed the validity of the update, they cast a vote on the blockchain, by invoking the governance contract.

Given a sufficient quorum of positive votes is reached on-chain, the contract can be deployed and upgraded during the CD phase. The timelock ensures that each stakeholder has time to react to such an imminent upgrade (for example, by exiting their funds before an upgrade is applied that would be to their disadvantage).

Recall that we assume a cooperative multi-party setting, e.g., with a number of different organisations cooperating on the basis of one or more smart contracts.
While this cooperation should be advantageous for each organisation with their respective business goals, the organisations in general have limited trust in one-another.
Still, given the cooperative setting, before a change to a smart contract is to be deployed, the organisations should ideally discuss about and agree on the change.
Nevertheless, given the limited mutual trust, none of the organisations should simply believe that another organisation truthfully implements the change.
Hence, our conceptual solution does not rely on any upfront consensus; instead, each organisation implements a set of tests in order to check that the smart contract code does not result in negative impact for themselves, including a broken integration with private systems of a given organisation.
These tests are run locally by each organisation, and only if the results are acceptable the organisation will vote in favour of the new version.

\subsection{Deterministic Registry and Deployment}
\label{sec:registry}
A core enabler of software builds that are verified in a distributed manner are deterministic builds (often also denoted as reproducible or attestable builds), where a valid source code version is guaranteed to generate identical builds across environments~\cite{jamthagen2016exploiting}.
A deterministic build is achieved by including all environment (e.g., compiler configuration, hardware setup) and other external factors influencing the build in the build version. This is easily achieved by containerisation of the build process (e.g., via \textit{Docker containers})~\cite{casalicchio2020container}.

In addition, an important consideration for deployments in a blockchain environment are deterministic deployments. The deployment of a smart contract involves assigning a blockchain address to it. If this process is deterministic w.r.t. the bytecode of the contract, it essentially allows to commit to a contract on-chain, without (yet) deploying it (or storing its bytecode).\footnote{On chains without CREATE2, bytecode hashes can be stored and compared directly, at the cost of an additional on-chain storage operation.} Such a feature was introduced to the Ethereum Virtual Machine (EVM) in EIP-1014~\cite{buterin2018eip1014}.

Next, without loss of generality, we generalise the address derivation function from EIP-1014 and describe the \textit{Deterministic Registry}, a variant of the registry pattern, which makes direct use of this functionality to enable deterministic deployments.

\subsubsection{Deterministic Address Derivation.}
EIP-1014 introduced the \verb|CREATE2| primitive, which uses a hash of the contracts initialisation code, a salt, the sender's address, and a fixed offset to produce a hash, which last 20 bytes are used as contract address~\cite{buterin2018eip1014}. 
Using the conventions of \cite{wood2014ethereum}, we abbreviate the address function for version $v_i$ as

\begin{equation}
v_i = \text{ADDR}(s, \text{KEC}(C_i), \zeta),
\end{equation}
where $i$ is the version identifier, $s$ the sender's address, \text{KEC} produces the Keccak-256 hash from this versions contract init code $C_i$, and $\zeta$ is the salt. For our purposes, we use $\zeta = i$, and set $s$ to the address of the registry, ensuring only it can deploy valid contracts. We, thus, write $v_i = \text{ADDR}(C_i, i)$ for short. However, a given version update will not only contain modifications to a singular contract. To support arbitrary modifications to the code base per version, one contract needs to implement the version controller pattern, which will include all contract addresses of the project in its bytecode, each, in turn, derived from \text{ADDR}. This is easily done, by e.g., making them part of its constructor arguments. Consequently, the version controller servers as anchor for any given version. Thus, we write 
\begin{equation}
v_i = \text{ADDR}(C_i(C_0,...,C_n), i),
\end{equation} for short. Where $v_i$ is the resulting contract address of the version controller contract for version $i$. Now, modifying any contract $C_0,...,C_n$ will change its corresponding address, propagating upwards and resulting in changing $v_i$.

\subsubsection{Deterministic Registry.}
Using the address derivation function, the registry can achieve the invariant \textit{version authenticity}: A contract in the upgrade proposal must be identical with the later deployed contract, to avoid the inclusion of malicious instructions post-voting.
Now, when the registry is invoked with the request to deploy a contract for update $i$, the registry uses its assigned governance contract to verify a proposal for $v_i$ exists, a quorum on it was reached, and the timelock has expired. It then verifies whether the derived address of the \emph{to-be-deployed} contract is identical with $v_i$ (in the proposal), and then proceeds to deploy and/or upgrade the contract, using the assigned proxy contracts. 

The deterministic registry is designed to be pluggable with different governance and proxy implementations.
Optionally, the registry can keep track of older versions to support gradual updates for e.g., canary or A/B testing~\cite[Ch. 5]{Bass2015DevOpsBook}.
\subsection{Continuous Integration \& Delivery}
\label{sec:CI}
In this Section, we will discuss the proposed CI and CD workflow in more detail. We will also outline a more automated variant, given some centralised workflow execution engine is available.
According to Bass et al.~\cite{Bass2015DevOpsBook}, \textit{continuous integration (CI)} stops after unit tests, whereas continuous delivery includes more elaborate tests, possibly including integration, performance, or (automated) user acceptance tests---but no automatic deployment. If the final quality gate is automated, i.e., sufficiently good results in the tests lead to deployment, they define it as \textit{continuous deployment (CD)}. We adopt their definitions, but for simplicity only differentiate CI and CD; 
the difference between CD and continuous delivery is not decisive here, though we acknowledge that in practice organisations should be precise about the scope of testing and the resulting implications.
%
%
\begin{figure}[tb]
    \centering
    \includegraphics[width=0.8\linewidth]{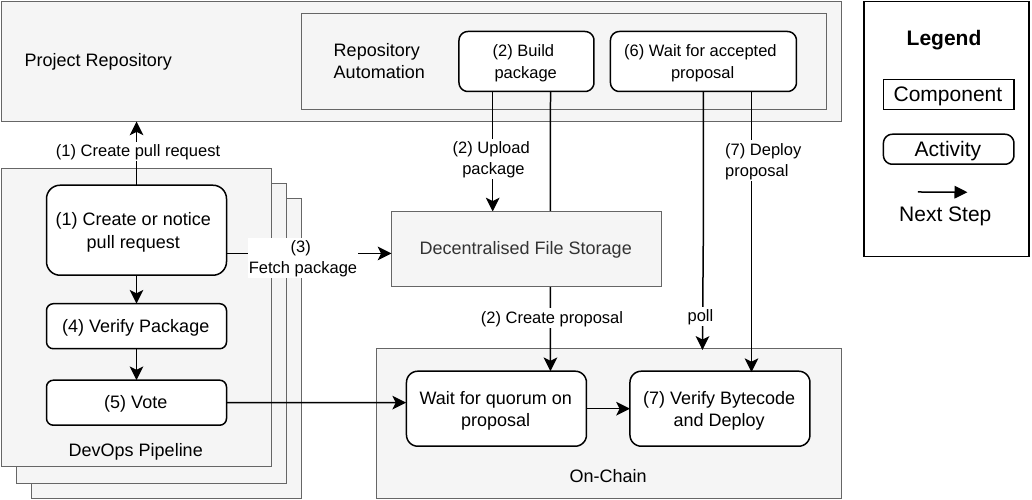}
    \caption{Overview of the CI/CD workflow with decentralised file storage integration.}
    \label{fig:workflow}
\end{figure}
\autoref{fig:workflow} depicts an overview of the CI/CD workflow.
In our proposal, CI works as follows.
When a pre-determined action in the centralised code version repository (e.g., new commit to the main branch) occurs, a new deterministic upgrade package is created from the repository files (See~\autoref{fig:workflow}, step (1) and (2)).

One of the core features of CI is automation.
Thus, we introduce a special \textit{package proposer}\footnote{This should not be confused with the timelock specific \textit{proposer} role of the OpenZeppelin Governor implementation. To avoid confusion, we adopt the terminology \textit{package proposer}.} role to allow autonomous scripts to access functions of the deterministic registry and governance contract. Using these functionalities, the proposal creation and execution can be automated. However, such (centralised) automation increases the reliance on the repository platform. Depending on the trust model, package build, proposal creation, and proposal execution can be handled by a stakeholder, who takes on the role of package proposer. To further reduce reliance on a (potentially) centralised repository platform, we suggest upgrade proposals to be additionally hosted on a decentralised file storage. However, this is not strictly necessary, given all stakeholders ensure storage locally.
We do not envision restrictions on who is able to trigger package proposals. However, a single proposer rule could be enforced and elected in a e.g., round-robin fashion. As a malicious single proposer could censor new updates, it is paramount to ensure the role can be re-assigned.
As additional optional step, automated centralised tests are executed on the repository platform.

The package proposer then creates a new proposal on the governance contract for the new version $i$, which includes the expected deterministically derived address $v_i$, and a reference to the proposal package (on the decentralised file storage).

All stakeholders observe this contract. Once the proposal including $v_i$ is available, each stakeholder fetches the package from the given reference, builds the contracts, verifies that $v_i = \text{ADDR}$, and executes all tests. In addition, each stakeholder may perform private tests not included in the central repository. This set should include tests that verify that the new version is not disadvantageous to them, and checks that the new version still integrates with their off-chain systems (see~\autoref{fig:workflow}, (3) and (4)). 

Once the run of a stakeholder's pipeline is complete, criteria are applied to determine whether the results are sufficiently good and the new version should be accepted (or rejected); accordingly, the stakeholder casts its vote on the proposal  (\autoref{fig:workflow}, (5)). The proposer polls the governance smart contract periodically to observe when the voting round is finished and whether the new version is accepted. If it is accepted, the proposer advances to the next build steps. Otherwise, the pipeline is stopped and the version rejected.
%
%
Continuous integration may trigger automated deployment, in which case the setup is an instance of a continuous deployment pipeline as per~\cite{Bass2015DevOpsBook}. Deployment may happen to public testnets before it is performed on the mainnet.
Mainnet deployment is expected to be triggered manually, which might include additional manual code audits~\cite{wohrer2021devops}. Such considerations are case specific and must be considered before casting a vote.
Given a pre-configured quorum on votes is reached on-chain, the contract can be deployed and upgraded during the CD phase.

A timelock delay ensures that each stakeholder has time to react to an imminent accepted proposal, and must elapse before a proposal is executed.

For deployment, a special role, the \textit{propagator} (which again facilitates automation) deploys the new version of the smart contract (\autoref{fig:workflow}, (7)) through the deterministic registry, by providing the version identifier $i$ and the bytecode. The registry verifies that a corresponding proposal on $i$ was accepted and that the bytecode would result in the expected address $v_i = \text{ADDR}$. Subsequently, it executes the upgrades via the proxy upgrade mechanism.
\subsection{Decentralised Governance}
\label{sec:gov}
\begin{table}[b]
\centering
\scriptsize
\caption{Stakeholder roles in the governance architecture.}
\label{tab:stakeholder_roles}
\begin{tabular}{p{1.5cm}p{5cm}p{.2cm}p{4cm}}
\toprule
\textbf{Role} & \textbf{Purpose} & &\textbf{Example Assignment} \\
\midrule
\verb|Stakeholder| & Full governance participation & & Human team member (e.g., lead developer, product owner) \\
\midrule
\verb|Package| \verb|Proposer| & Least-privilege role for automating submitting package proposals & & Automated CI/CD script \\
\midrule
\verb|Propagator| & Least-privilege role for automating queuing and execution of successful proposals & & Automated monitor script \\
\midrule
\verb|Voter| & Casts for/against votes  & & Human stakeholder assigned to review the proposal \\
\bottomrule
\end{tabular}
\end{table}
The governance contract enforces three parameters for each proposal vote: (i) a voting delay, an amount of time that has to elapse before the votes are tallied, allowing sufficient time to participate, (ii) a voting period, which defines the duration the vote is active, which importantly also serves as an upper bound ensuring the protocol cannot be deadlocked, and (iii) a timelock delay, which comprises a delay that needs to elapse before the agreed upon version is deployed, so all users can react and prepare accordingly. Each of these parameters can be changed, by casting a dedicated proposal, which undergo the same voting process as an update proposal.
In addition to handling package proposals, governance also involves assigning identities (blockchain account addresses) to pre-defined roles. An overview of the proposed roles is given in \autoref{tab:stakeholder_roles}.
We do not further discuss special cases of voting, such as delegated voting. At its most complex, the Governance vote can be weighted based on the (delegated) ownership of a dedicated governance token. See \cite{han2025daoreview} for an overview.

As the identity of involved stakeholders may change over time, it defines procedures to nominate new and release existing stakeholders. In summary, the following proposals exist:
(i) \textit{Upgrade Proposal}: Proposes a new version for deployment; (ii) \textit{Stakeholder Change}: Nominates a new stakeholder or releases an existing one; (iii) \textit{Role Assignment}: Grants or revokes one of the specialised roles; and (iv) \textit{Parameter Change}: Modifies the parameters voting delay, voting period, or timelock delay.

\section{Evaluation}
\label{sec:eval}

Our contribution comprises two artifacts, (i) the reference architecture and (ii) the deterministic registry pattern. 
To evaluate these contributions throughly, three properties need to be demonstrated.
First, \textbf{feasibility}: the architecture and the deterministic registry pattern must be realisable as an integrated, working prototype. To this end, we provide a reference implementation, openly available together with a replication package, and assess the feasibility against actual updates of a real-world blockchain application (\autoref{ssec:impl}).
Second, \textbf{correctness}: the implementation must enforce the workflow defined by the architecture--in particular, the ordering constraints of the deployment lifecycle (e.g., no deployment without a prior successful vote) and the version authenticity (bytecode-integrity) invariant of the deterministic registry. We verify this through process-conformance checking: we mine event logs emitted by the on-chain contracts and replay them against the expected process model, covering both conforming and non-conforming execution paths (\autoref{ssec:correctness}). This technique has been shown to be able to discover unintended behaviour of blockchain-based implementations, such as security relevant bugs~\cite{hobeck2024suitability}.
Third, \textbf{cost}: the cost overhead must be quantifiable so that practitioners can assess economic viability. We study gas costs incurred for each workflow step of our implementation (\autoref{ssec:cost}).
Subsequently, in \autoref{sec:disc} we discuss the results and revisit R1--R3.

\subsection{Implementation \& Feasibility Assessment}\label{ssec:impl}
We implemented the architecture described in \autoref{sec:solution}, consisting of three integrated components:
(i)~on-chain contracts,
(ii)~an off-chain containerised build environment, and
(iii)~a GUI, serving as dashboard for stakeholders.
Our implementation is available in form of a replication package, including our evaluation and scripts and documentation to reproduce it.\footnote{\label{fn:rep}Repository is available at \url{https://github.com/ISDO-TUM/dobbie}. An archived version is available at \url{https://doi.org/10.5281/zenodo.20286777}.}
We implement the contracts (i) in Solidity, and use the OpenZeppelin library, specifically, their Governance and BeaconProxy implementation.\footnote{\url{https://docs.openzeppelin.com/contracts/5.x/governance} and \url{https://docs.openzeppelin.com/contracts/5.x/api/proxy\#beaconproxy}, both accessed 2026-03-24.}
To bootstrap the system, we also implement a contract factory.
The off-chain environment (ii) performs three checks on each package proposal: (a)~\emph{integrity}---recomputing $\text{ADDR}$ from the packaged bytecode and asserting they match the on-chain proposal~$v_i$;
(b)~\emph{standard tests}---building a Docker container\footnote{\url{https://www.docker.com}, accessed 2026-03-24.} from the package and running the project's test suite in a reproducible environment;
and (c)~\emph{private tests}---injecting stakeholder-specific private tests into the container.
Results are persisted locally and exposed via a REST API to be consumed by a locally-running GUI.
The GUI (iii) allows stakeholders to monitor proposals, inspect verification results, and cast votes.
It reads state directly from the blockchain and presents proposal lifecycles, voting progress, and package metadata.

To assess \textbf{feasibility}, we conducted an experiment involving deploying and upgrading a real-world Ethereum application across two geographic regions.
We chose the \textit{Hub-Portal-Chat} open source application, with the goal to represent a common multi-contract dApp architecture used in a production environment. 
Hub-Portal-Chat is a decentralised social media app, operated on six compatible blockchain networks.\footnote{See \url{https://hub-portal-chat.vercel.app/} and \url{https://github.com/Mystique85/Hub-Portal-Chat}, for the open source repository. Both accessed 2026-03-27.}
Its structure, which includes a central hub coordinating multiple peripheral contracts, fits our version controller pattern.
We used two automated workflow scripts, hosted on a GitHub repository we set up for the project.\footnote{\label{fn:rep2}The repository is available at \url{https://github.com/kirillinoz/dobbie-template-chat}. And an archived version (only source code) at \url{https://doi.org/10.5281/zenodo.20286850}.}
The workflow scripts automatically create a proposal as part of a merge request, and inject necessary meta data into the merge request text body. 
The simulation involved three distinct entities to test the distributed nature of the framework. 
We deployed two stakeholders, who act as authorised voters, and set the quorum to two approvals. One stakeholder was deployed on \textit{Amazon Web Services} in the \verb|eu-north-1c| zone, 
while the other stakeholder was operated in our local environment located in Munich.
Both environments were configured with access to the shared GitHub repository and distinct private test suites. As on-chain environment, we used the \verb|Sepolia| Ethereum test network.

In the first test case, a software update was pushed. The automated agent successfully proposed the package. 
The CI/CD pipelines of both stakeholders were successful, both voted for approval, the integrity check was positive, yielding a successful contract update.
A second scenario was simulated where one stakeholder rejected a proposed update. Consequently, the proposal failed to reach the required quorum, and the voting period ended.

\subsection{Correctness Benchmark}\label{ssec:correctness}
Our on-chain contracts emit events with each state transition: on proposal submission, during voting, queuing a proposal in the timelock, execution of a proposal, and when the deployment of a new version is  finalised. This serves as an immutable audit trace. We here use it to assess the correctness of our proposal.

To this end, we divide our benchmark into conforming and non-conforming (invalid execution) scenarios. Our CI/CD workflow (c.f.,~\autoref{fig:workflow}) implicitly defines a workflow model. From it, we derived a Petri Net that any given valid execution order must follow. 
Valid executions have three main variants: success, rejection, and timeout; these were covered in the manually specified reference process model.
Against this process model, we compare each trace (sequence of events) in the logs that were generated by executing each scenario. Conformance of the trace against the model was verified by applying the \textit{Token-Based Replay} algorithm~\cite[Ch.~7.2]{vanderaalstProcessMiningDiscovery2011}, using the \verb|pm4py| process mining library~\cite{pm4py}. A fitness score of $100\%$ indicates no deviation between the observed execution and the expected process model, which we achieved for all conforming logs. For non-conforming scenarios, we confirmed that each contained a reverted transaction. 
%
%
\subsubsection{Scenarios.}
\autoref{tab:scenarios} gives an overview of the scenarios we executed. The Petri net defines three conforming scenarios, that cover all valid paths through the workflow. Based on this, we define seven non-conforming scenarios, simulating possible violations against the enforced order of events. For the scenarios, we verify our upgrade proposal implementation, as the other proposal flows are equal to OpenZeppelin's \verb|Governor| (version 5) implementation.
All scenarios were executed on a local Hardhat blockchain simulation environment with three registered stakeholders (equal voting weight), a voting delay of 1~block ($\approx$10\,s), a voting period of 2\,160~blocks ($\approx$6\,h at 10\,s/block), a quorum of two out of three stakeholders, and a timelock delay of 7\,200\,s (2\,h).

\begin{table}[t]
\centering
\scriptsize
\caption{Scenarios for the correctness benchmark}
\label{tab:scenarios}
\begin{tabular}{rp{.2cm}l}
\toprule
\textbf{Scenario} & & \textbf{Description} \\
\midrule
C1 & & Valid upgrade proposal with successful quorum  \\
C2 & & Valid upgrade proposal with rejected (majority of against votes) quorum \\
C3 & & Valid upgrade proposal with rejected (no majority reached) quorum \\
\midrule
N1 & & Premature proposal execution (without queuing it in the timelock) \\
N2 & & Premature proposal execution (without an elapsed timelock delay) \\
N3 & & Late vote (after the voting period) \\
N4 & & Early vote (before the voting delay) \\
N5 & & Double vote \\
N6 & & Premature queuing of the proposal (before a quorum was reached) \\
N7 & & Tampered proposal (submitting a tampered contract address) \\
\bottomrule
\end{tabular}
\end{table}
\begin{figure}[t]
    \centering
    \includegraphics[width=1\linewidth]{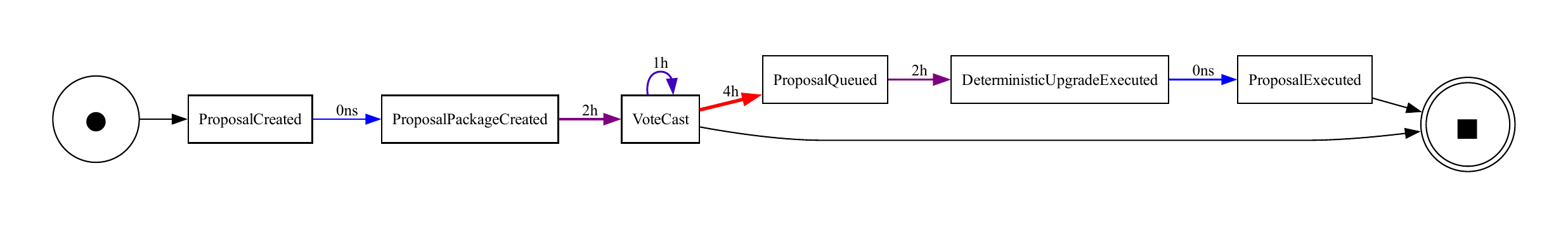}
    \vspace{-2em}
    \caption{Performance-annotated graph mined from conforming scenario logs (C1--C3).}
    \label{fig:pm}
\end{figure}
Our benchmark confirmed the correct implementation of every reachable path and that non-conforming execution attempts  result in reverted transactions. Notably, with N7, we verify that the deterministic registry rejects a deployment where the submitted bytecode and expected address mismatch.

To confirm the effect of the timelock, voting delay, and voting period, we generated a performance-annotated directly-follows graph~\cite[Ch.~11.4.]{dumasFundamentalsBusinessProcess2018}, by mining the event logs received from executing the conforming scenarios (C1--C3), using \verb|pm4py|~\cite{pm4py}. Using it, we confirm that all discovered delays in the log are explainable given our setup and configuration.

\autoref{fig:pm} shows the resulting graph. Boxes correspond to the emitted events during an execution. 
Edges show observed sequences of events.
Edges are labelled with mean transition times across all traces containing that edge.
We observe $0$\,ns between \verb|ProposalCreated| and \verb|ProposalPackageCreated|, as they are co-emitted in the same transaction. $2$\,h from proposal creation to the first vote \verb|VoteCast|, which is the mean of our simulated vote-arrival times across scenarios ($1$\,h for C1, $2$\,h for C2, $3$\,h for C3). A $1$\,h delay between subsequent votes, which correspond to our simulation of the second vote. Unsuccessful scenarios (C2 and C3) do not proceed further with deployment, which corresponds with one edge leading to the end of the graph. For C1, we observe a $4$\,h delay, which is the remaining voting period after the first vote, followed by the $2$\,h configured timelock delay.
\verb|DeterministicUpgradeExecuted| and \verb|ProposalExecuted| are again co-emitted in the same transaction.
\begin{wraptable}{r}{0.4\textwidth}
\centering
\scriptsize
\vspace{-.5em}
\caption{Gas cost evaluation results.}
\label{tab:gas_costs}
\begin{tabular}{l r}
\toprule
\textbf{Governance Step} & \textbf{Gas Used} \\
\midrule
Grant Role (one-time) & 51\,311 \\
Propose Package       & 115\,306 \\
Cast Vote (first)     & 75\,463 \\
Cast Vote (subsequent)& 58\,363 \\
Queue Proposal        & 132\,671 \\
Execute Proposal      & 74\,352 \\
\midrule
\textbf{Total Workflow} & \textbf{507\,466} \\
\bottomrule
\end{tabular}
\vspace{-5em}
\end{wraptable}
\subsection{Gas Cost Analysis}\label{ssec:cost}
As gas costs on the EVM are deterministic, they can be reliably measured via local simulation. We simulated a complete workflow lifecycle on a local Hardhat instance (Solidity~\verb|0.8.28|, optimizer enabled), using the same configuration as in the correctness benchmark. 
\autoref{tab:gas_costs} presents the per-step gas consumption for a complete lifecycle. Queuing is the most expensive step ($\approx$133k gas), as it requires storage of the queued operation. The first vote costs more than subsequent votes ($\approx$75k vs.\ $\approx$58k gas) due to EVM cold storage slot initialisation~\cite{wood2014ethereum}. Proposal creation ($\approx$115k gas) reflects the on-chain storage of the IPFS identifier and governance calldata.
If costs were calculated at a gas price of $20$\,gwei and taking the 30-day average (25~February--26~March~2026) ETH price of \$\,2{,}073\footnote{Data from \url{https://www.coingecko.com}, accessed 2026-03-26.} a full lifecycle would amount to $\$\,21$ on the Ethereum mainnet. 

%
%
\section{Discussion \& Conclusion}\label{sec:disc}

The \textbf{requirements R1-R3} were listed in \autoref{tab:requirements}, all of which drove the design of our solution. R1 necessitates individual testing for each participant, which we tested with positive outcome in the feasibility assessment, \autoref{ssec:impl}, for both approval and rejection outcomes and with geographic distribution.
R2 expresses the need for consensus and non-repudiation, which we asserted with the correctness benchmark in \autoref{ssec:correctness}. C1-C3 confirm all three valid termination paths, and N1-N6 rejected relevant deviations from the permitted execution path. Non-repudiation is given with the immutable on-chain history, particularly the event log. Note, that our current implementation does not emit events for rejected proposals--an implementation choice that could be changed easily if so desired.
R3, version authenticity, is achieved by the deterministic registry pattern and tested specifically with N7: submitting tampered bytecode is rejected, due to the cryptographic binding between code voted on and permitted update address.

A practical concern is \textbf{cost}: \$\,21 per update is relatively high, and might be impractical for numerous applications, particularly when updates are frequent. This observation aligns with prior findings on layer one cost for workflow execution~\cite{stiehle2024cost}.
The cost burden can be alleviated by using layer two or third-generation blockchains, which have been shown to reduce cost significantly~\cite{stiehle2024cost}.
Since EIP-4844~\cite{eip4844}, transaction costs on optimistic rollups have decreased substantially~\cite{galaxy2024dencun}. On the Optimism network, execution gas prices of approx. $0.001$\,gwei have been observed,\footnote{\url{https://chainspect.app/chain/optimism}, accessed 2026-03-26.}
compared to the 20\,gwei we assumed for our mainnet cost calculation. 
At these rates, the execution cost would amount to less than \$\,0.01, with per-transaction mainnet data posting fees adding marginal overhead. 
At such cost levels, even a per-pull-request governance model becomes economically viable. However, long-term studies are required. In principle, predicting costs for blockchain applications is challenging, as it relies on network factors, e.g., congestion, or economic factors, e.g., exchange rates~\cite{stiehle2024cost}.

In terms of \textbf{limitations}, our approach relies on a blockchain notion of time, specifically where time is approximated as block-time, which can vary slightly -- within protocol bounds, validator discretion might extend to seconds~\cite{ladleifTimeBlockchainBasedProcess2020}.
Our approach relies on time for voting delay, voting periods, and timelock periods.
If these parameters are on the order of minutes or hours, as would be common, the implication is negligible; for very short durations, the security margins degrade.

As future work, the governance workflow could alternatively be performed off-chain, e.g., in state channels~\cite{stiehle2023processchannels}, settling only the final outcome on-chain -- further reducing cost at the expense of additional liveness assumptions. 

To conclude, we presented an integrated architecture for decentralised deployment of blockchain applications, bridging governance, upgradeability, and DevOps practices. The deterministic registry pattern provides cryptographic binding between voted-on bytecode and deployed addresses. Our open-source implementation and replication package are available to support adoption and reproducibility.
\vspace{-1.5em}
\subsubsection*{Data Availability.} All our artefacts are available as disclosed in~\autoref{sec:eval}, including our evaluation and scripts and documentation to reproduce it. Specifically, see Footnote~\ref{fn:rep} and Footnote~\ref{fn:rep2}.
%
%
%
\bibliographystyle{splncs04}
\bibliography{bib}

@inproceedings{bodell2023proxy,
  title={Proxy hunting: Understanding and characterizing proxy-based upgradeable smart contracts in blockchains},
  author={Bodell III, William E and Meisami, Sajad and Duan, Yue},
  booktitle={USENIX Security},
  pages={1829--1846},
  year={2023}
}

@misc{eip2535,
  author       = {Nick Mudge},
  title        = {{ERC-2535: Diamonds, Multi-Facet Proxy}},
  howpublished = {\url{https://eips.ethereum.org/EIPS/eip-2535}},
  note         = {Ethereum Improvement Proposals, no. 2535, Online},
  month        = feb,
  year         = {2020}
}

@article{nasr2024hybrid,
  title={A Hybrid Framework to Implement DevOps Practices on Blockchain Applications (DevChainOps).},
  author={Nasr, Ramadan and Marie, Mohamed I and El Sayed, Ahmed},
  journal={IJACSA},
  volume={15},
  number={6},
  year={2024}
}

@inproceedings{wohrer2021devops,
  title={{DevOps} for {Ethereum} Blockchain Smart Contracts},
  author={W{\"o}hrer, Maximilian and Zdun, Uwe},
  booktitle={2021 IEEE International Conference on Blockchain (Blockchain)},
  pages={244--251},
  year={2021},
  organization={IEEE}
}

@book{Bass2015DevOpsBook,
  author = {Bass, Len and Weber, Ingo and Zhu, Liming},
  title = {DevOps: A Software Architect's Perspective},
  year = {2015},
  publisher = {Addison-Wesley Professional},
  isbn = {978-0-134-04984-7},
  cdoi = {10.1007/978-0-134-04984-7},
}

@book{vanderaalstProcessMiningDiscovery2011,
  title = {Process {{Mining}}: {{Discovery}}, {{Conformance}} and {{Enhancement}} of {{Business Processes}}},
  shorttitle = {Process {{Mining}}},
  author = {van der Aalst, Wil M. P.},
  options = {useprefix=true},
  year = {2011},
  publisher = {{Springer, Heidelberg}},
  clocation = {{Berlin, Heidelberg}},
  ccdoi = {10.1007/978-3-642-19345-3},
  ccurl = {http://link.springer.com/10.1007/978-3-642-19345-3},
  curldate = {2022-08-22},
  isbn = {978-3-642-19344-6 978-3-642-19345-3},
  langid = {english}
}

@book{dumasFundamentalsBusinessProcess2018,
  title = {Fundamentals of {{Business Process Management}}},
  author = {Dumas, Marlon and La Rosa, Marcello and Mendling, Jan and Reijers, Hajo A.},
  year = {2018},
  publisher = {{Springer, Heidelberg}},
  cdoi = {10.1007/978-3-662-56509-4},
  isbn = {978-3-662-56508-7 978-3-662-56509-4},
  langid = {english}
}

@inproceedings{ladleifTimeBlockchainBasedProcess2020,
  title = {Time in {{Blockchain-Based Process Execution}}},
  booktitle = {EDOC},
  cbooktitle = {2020 {{IEEE}} 24th {{International Enterprise Distributed Object Computing Conference}} ({{EDOC}})},
  author = {Ladleif, Jan and Weske, Mathias},
  year = {2020},
  ccmonth = oct,
  pages = {217--226},
  publisher = {{IEEE}},
  ccdoi = {10.1109/EDOC49727.2020.00034},
  cisbn = {978-1-72816-473-1}
}

@InProceedings{stiehle2023processchannels,
author="Stiehle, Fabian
and Weber, Ingo",
ceditor="Di Francescomarino, Chiara
and Burattin, Andrea
and Janiesch, Christian
and Sadiq, Shazia",
title="Process Channels: A New Layer for Process Enactment Based on Blockchain State Channels",
booktitle="BPM",
year="2023",
publisher="Springer",
caddress="Cham",
pages="198--215",

}

@inproceedings{stiehle2022blockchain,
  title={Blockchain for business process enactment: a taxonomy and systematic literature review},
  author={Stiehle, Fabian and Weber, Ingo},
  booktitle={BPM: Forum},
  pages={5--20},
  year={2022},
  series={LNBIP},
  corganization={Springer},
  volume = {459}
}

@misc{buterin2018eip1014,
  author       = {Vitalik Buterin},
  title        = {{EIP-1014: Skinny CREATE2}},
  howpublished = {\url{https://eips.ethereum.org/EIPS/eip-1014}},
  note         = {Ethereum Improvement Proposal},
  number       = {1014},
  year         = {2018},
  month        = apr
}

@article{hobeck2024suitability,
  title={On the Suitability of Process Mining for Enhancing Transparency of Blockchain Applications},
  author={Hobeck, Richard and Klinkm{\"u}ller, Christopher and Bandara, HMN Dilum and Weber, Ingo and van der Aalst, Wil},
  journal={Business \& Information Systems Engineering},
  pages={1--20},
  year={2024},
  publisher={Springer}
}

@inproceedings{stiehle2024cost,
  title={The Cost of Executing Business Processes on Next-Generation Blockchains: The Case of Algorand},
  author={Stiehle, Fabian and Weber, Ingo},
  booktitle={International Conference on Business Process Management},
  pages={89--105},
  year={2024},
  organization={Springer}
}

@article{pm4py,  
title = {{PM4Py}: A process mining library for Python},  
journal = {Software Impacts},  
volume = {17},  
pages = {100556},  
year = {2023},  
issn = {2665-9638},  
cdoi = {https://doi.org/10.1016/j.simpa.2023.100556},  
curl = {https://www.sciencedirect.com/science/article/pii/S2665963823000933},  
author = {Alessandro Berti and Sebastiaan van Zelst and Daniel Schuster},  
}

@article{casalicchio2020container,
  title={The state-of-the-art in container technologies: Application, orchestration and security},
  author={Casalicchio, Emiliano and Iannucci, Stefano},
  journal={Concurrency and Computation: Practice and Experience},
  volume={32},
  number={17},
  pages={e5668},
  year={2020},
  publisher={Wiley Online Library}
}

@misc{wood2014ethereum,
  title={Ethereum: A secure decentralised generalised transaction ledger},
  author={Wood, Gavin and others},
  howpublished={Ethereum project yellow paper},
  year={2014}
}

@article{shahin2017continuous,
  title={Continuous integration, delivery and deployment: a systematic review on approaches, tools, challenges and practices},
  author={Shahin, Mojtaba and Babar, Muhammad Ali and Zhu, Liming},
  journal={IEEE access},
  volume={5},
  pages={3909--3943},
  year={2017},
  publisher={IEEE}
}

@article{gorski2021towards,
  title={Towards Continuous Deployment for Blockchain},
  author={G{\'o}rski, Tomasz},
  journal={Applied Sciences},
  volume={11},
  number={24},
  pages={11745},
  year={2021},
  publisher={MDPI}
}

@article{gorski2021continuous,
  title={Continuous Delivery of Blockchain Distributed Applications},
  author={G{\'o}rski, Tomasz},
  journal={Sensors},
  volume={22},
  number={1},
  pages={128},
  year={2021},
  publisher={MDPI}
}

@inproceedings{bankar2021blockchain,
  title={Blockchain based framework for Software Development using {DevOps}},
  author={Bankar, Sandip and Shah, Deven},
  booktitle={Intl. Conf. Nascent Technologies in Engineering (ICNTE)},
  pages={1--6},
  year={2021},
  organization={IEEE}
}

@inproceedings{akbar2022toward,
  title={Toward Effective and Efficient {DevOps} using Blockchain},
  author={Akbar, Muhammad Azeem and Mahmood, Sajjad and Siemon, Dominik},
  booktitle={EASE},
  pages={421--427},
  year={2022}
}

@inproceedings{beller2019blockchain,
  title={Blockchain-based software engineering},
  author={Beller, Moritz and Hejderup, Joseph},
  booktitle={ICSE-NIER},
  pages={53--56},
  year={2019},
  organization={IEEE}
}

@article{han2025daoreview,
  title={A review of DAO governance: Recent literature and emerging trends},
  author={Han, Jungsuk and Lee, Jongsub and Li, Tao},
  journal={Journal of Corporate Finance},
  volume={91},
  pages={102734},
  year={2025},
  publisher={Elsevier}
}

@inproceedings{benedetti2024comparative,
  title={A comparative gas cost analysis of proxy and diamond patterns in EVM blockchains for trusted smart contract engineering},
  author={Benedetti, Anto and Henry, Tiphaine and Tucci-Piergiovanni, Sara},
  booktitle={Intl. Conf. Financial Cryptography and Data Security},
  pages={207--221},
  year={2024},
  organization={Springer}
}

@article{kumar2025leagan,
  title={LEAGAN: A Decentralized Version-Control Framework for Upgradeable Smart Contracts},
  author={Kumar, Gulshan and Saha, Rahul and Conti, Mauro and Buchanan, William J},
  journal={IEEE TSC},
  year={2025},
  publisher={IEEE}
}

@incollection{xu2025patternextended,
  title={An extended pattern collection for blockchain-based applications},
  author={Xu, Xiwei and Pautasso, Cesare and Lo, Sin Kuang and Zhu, Liming and Lu, Qinghua and Weber, Ingo},
  booktitle={Transactions on Pattern Languages of Programming V},
  pages={67--117},
  year={2025},
  publisher={Springer}
}

@inproceedings{xu2018pattern,
  title={A pattern collection for blockchain-based applications},
  author={Xu, Xiwei and Pautasso, Cesare and Zhu, Liming and Lu, Qinghua and Weber, Ingo},
  booktitle={EuroPLoP},
  pages={1--20},
  year={2018}
}

@inproceedings{jamthagen2016exploiting,
  title={Exploiting trust in deterministic builds},
  author={J{\"a}mthagen, Christopher and Lantz, Patrik and Hell, Martin},
  booktitle={SAFECOMP},
  pages={238--249},
  year={2016},
  organization={Springer}
}

@inproceedings{saleh2024blockchain,
  title={Blockchain for securing CI/CD pipeline: A review on tools, frameworks, and challenges},
  author={Saleh, Sabbir M and Madhavji, Nazim and Steinbacher, John},
  booktitle={2024 7th Conference on Cloud and Internet of Things (CIoT)},
  pages={1--5},
  year={2024},
  organization={IEEE}
}

@misc{eip4844,
  title        = {{EIP}-4844: Shard Blob Transactions},
  author       = {Buterin, Vitalik and Dankrad, Feist and Diederik, Loerakker and Lightclient and Rory, Skinner},
  year         = {2022},
  howpublished = {Ethereum Improvement Proposals},
  url          = {https://eips.ethereum.org/EIPS/eip-4844},
  note         = {Accessed: 2026-03-26}
}

@misc{galaxy2024dencun,
  title        = {150 Days After Dencun},
  author       = {Pokorny, Zack},
  year         = {2024},
  month        = {August},
  howpublished = {Galaxy Research},
  url          = {https://www.galaxy.com/insights/research/ethereum-150-days-after-dencun},
  note         = {Accessed: 2026-03-26}
}

\end{document}